# GLOBULAR CLUSTER SYSTEMS OF EARLY-TYPE GALAXIES

Sidney van den Bergh

Dominion Astrophysical Observatory

National Research Council

5071 West Saanich Road

Victoria, British Columbia, V8X 4M6

Canada

E-mail: Sidney.vandenbergh@hia.nrc.ca

## ABSTRACT

Properties of 53 globular cluster systems are investigated. Strong correlations are found between parent galaxy luminosity and both the slope of the radial density profile for clusters and the width of the cluster color (metallicity) distribution. These correlations are in the sense that the most luminous early-type galaxies are embedded in cluster systems that have the shallowest radial gradients and exhibit the broadest color distributions. The data suggest a scenario in which luminous early-type galaxies have a more complex evolutionary history than fainter ones. A problem with the interpretation of the present data is that it is difficult (or impossible) to disentangle the strongly correlated effects of high parent galaxy luminosity, presence of a core or boxy isophotes, and shallow radial cluster density gradients.

Subject headings: galaxies: clusters

**1.    INTRODUCTION**

The present morphology of a galaxy depends on the physical nature of its ancestral object(s). It is now generally believed (e.g. Kissler-Patig 1997a) that early-type galaxies of the "core" sub-type were likely assembled by mergers between mainly stellar ancestral objects. On the other hand early-type "power-law" galaxies are thought to have evolved by the collapse of a single protogalaxy, or from mergers between gas-rich ancestors; see van den Bergh (1998) for a review. It is therefore clearly of interest to search for possible correlations between the profile types of E, S0 and Sa parent galaxies and characteristics of their entourage of globular clusters. However, the interpretation of such correlations is rendered somewhat uncertain by the fact that power-law galaxies are, in the mean, much less luminous than galaxies having cores (Bender et al. 1989). It is therefore difficult to distinguish between those correlations that are mostly due to luminosity effects, and ones that are mainly driven by some aspect of evolutionary history that affects profile type. An additional complication is that cluster systems evolve more rapidly (mostly by destruction of low-mass clusters) in dense low-luminosity galaxies, than they do in lower density giant systems (Murali & Weinberg 1997).

## 2. PROPERTIES OF GLOBULAR CLUSTER SYSTEMS

The properties of globular cluster systems surrounding galaxies have been reviewed by Harris (1991). A recent compilation of data on cluster systems is given by Harris (1996). An up-to-date listing of some of the characteristics of globular cluster systems surrounding early-type galaxies is given in Table 1. This table is mainly based on a recent compilation by Kissler-Patig (1997b), that was supplemented by information on the radial luminosity distributions of parent galaxies by Faber et al. (1997), by Forbes (1997), and by Jaffe et al. (1994). In Table 1 galaxies with central cores in their radial profiles are denoted by "C", and those exhibiting power-law profiles by "P". (These are Types I and II of Jaffe et al., respectively). Also given in Table 1 are estimates of the specific globular cluster frequency S (Harris & van den Bergh 1981), the slope of the radial density distribution of clusters $\alpha$ [defined by $\sigma(cl) \propto r^{\alpha}$], and a characterization of the color (metallicity) distributions of the globular cluster systems. In the table "s" denotes a system with a relatively narrow color distribution, and "b" a cluster system that has a broad (or bimodal) color distribution. Also given are the normalized Fourier coefficients $\underline{a}(4)/\underline{a}$ describing the isophote shapes for parent galaxies. Positive and negative values of $\underline{a}(4)/\underline{a}$ denote "disky" and "bulgy" objects, respectively. Hubble classification types have <u>not</u> been included in Table 1 because E and S0 classification types by expert morphologists are only loosely correlated with their photometric characteristics (van den Bergh 1989). Unfortunately Table 1 contains only 53 entries. Most conclusions about the inter-relationships between various parameters describing cluster systems surrounding early-type galaxies will therefore remain highly tentative until more cluster observations become available.

## 3.    CORRELATIONS AMONG CLUSTER AND GALAXY PROPERTIES

The data in Table 1 can be used to establish three kinds of correlations: (1) those between different characteristics of globular cluster systems, (2) those between the properties of their parent galaxies, and (3) correlations between the characteristics of cluster systems and those of their parent galaxies. For such statistical discussions it is convenient to divide cluster systems into cluster-rich systems with $S > 5.0$, and cluster-poor ones with $S < 5.0$. Furthermore cluster systems can be divided into those exhibiting flat radial density distributions ($\alpha > -1.75$), and ones with steep radial density gradients ($\alpha < -1.75$). The parent galaxies can also be divided into luminous objects with $M_V < -21.5$ ($H_o = 75$ km sec$^{-1}$ Mpc$^{-1}$ assumed), and fainter ones with $M_V > -21.5$. Finally the parent galaxies can be assigned to boxy [$\underline{a}(4)$ = negative] or disky [$\underline{a}(4)$ = positive] sub-types.

### 3.1 Characteristics of globular cluster systems

Table 1 can be used to investigate the correlations between the specific globular cluster frequency S and the radial cluster density gradient $\alpha$, between S and the color (metallicity) distribution of clusters, and between $\alpha$ and the color distributions of clusters.

Figure 1 and table 2 show that systems with steep radial gradients tend to be cluster-poor with S < 5, whereas those with shallow radial gradients mostly have S > 5. A Chi-squared test of this 2x2 contingency table (Conover 1980) shows that there is only a 4% probability ($\chi^2 = 4.2$ with one degree of freedom) that specific cluster frequency and radial profile type are uncorrelated.

Table 3 shows the relation between the b (broad/binary) versus s (single/sharp) color distribution of clusters and the radial density gradients $\alpha$ of cluster systems. This 2x2 contingency table shows that there is less than a 0.1% probability ($\chi^2 = 14$, one d.f.) that the radial density gradients and the color distributions of clusters are uncorrelated. Perhaps unexpectedly, the color distributions of globular cluster systems do <u>not</u> correlate significantly ($\chi^2 = 1.7$ with one d.f.) with the specific cluster frequency S.

### 3.2 Relation between galaxy and cluster characteristics

Table 4 shows the values of Chi-squared for 2x2 contingency tables relating the radial profile type, absolute magnitude and isophote shape of parent galaxies with the specific frequency, radial gradient and color distribution of their globular cluster systems. The table shows that the most significant correlation observed is that between parent galaxy luminosity and the radial density gradient of cluster systems. This correlation is in the sense that the most luminous parent galaxies are surrounded by cluster systems having the shallowest gradients. The existence of such a relation was first established by Harris (1986, 1993). The only other significant correlation is that between parent galaxy luminosity and cluster color distribution. Luminous parent galaxies tend to have b (broad/binary) color distributions, hinting at a complex evolutionary history, whereas cluster systems of the s type, which have a single sharp peak in their color (metallicity) distributions tend to have lower luminosity parent galaxies.

### 3.3 Correlations between parent galaxy characteristics

Table 5 show a 2x2 contingency table between $\underline{a}(4)$ and the radial density profile type of the parent galaxy. This correlation, which has a probability of < 0.5% ($\chi^2 = 8.1$ with 1 d.f.) of being due to chance, is in the sense that boxy galaxies mostly have cores, whereas the majority of disky galaxies exhibit power-law profiles. For lower luminosity galaxies deviations from this relation may be produced by tidal distortions of the outer isophotes of early-type galaxies (Nieto & Bender 1989). Table 6 demonstrates that galaxies with cores are more luminous than those with power-law profiles ($\chi^2 = 7.5$ for one d.f.). Finally Figure shows a plot of the isophote shape parameter $\underline{a}(4)/\underline{a}$ versus galaxy luminosity $M_V$. This figure and Table show that the majority of "disky" early-type galaxies are fainter than $M_V = -21.5$ and that most "boxy" early-type galaxies are brighter than $M_V = -21.5$. The data in Table 7 show a significant correlation ($\chi^2 = 6.6$ for one d.f.) between the isophote shape parameter $\underline{a}(4)$ and parent galaxy luminosity. As has been known for many years (Bender et al. 1989) this table shows that early-type galaxies with boxy isophotes tend to be more luminous than disky ones.

## 4. SUMMARY AND CONCLUSIONS

The specific frequency of globular clusters in late-type galaxies is low. As a result it is not yet possible to draw any statistically significant conclusions about the global characteristics of the globular cluster systems associated with galaxies of types Sb, Sc and Ir. However, the situation is more favorable for the, generally richer, cluster systems that are associated with early-type galaxies. Table 1 summarizes the data that are presently available on 53 globular cluster systems associated with galaxies of types E, S0 and Sa. From this compilation it is seen that luminous parent galaxies, which usually have cores and boxy isophotes, are generally surrounded by globular cluster systems exhibiting shallow radial density gradients. Furthermore the globular clusters associated with luminous parent galaxies tend to have a broader color (metallicity) distribution than do those that are associated with less luminous parent galaxies.

It is a pleasure to thank Duncan Forbes for providing information on the radial profiles of some early-type galaxies.

# TABLE 1

PROPERTIES OF GLOBULAR CLUSTER SYSTEMS HAVING EARLY-TYPE PARENT GALAXIES[a]

| Designation | $M_v$ | Profile type | $100\underline{a}(4)/\underline{a}$ | S | α | Color |
|---|---|---|---|---|---|---|
| NGC 221 | -16.3 | P[b] | ... | 0.8 | 0.0 | ... |
| NGC 524 | -21.9 | C | ... | 4.8 | -1.71 | ... |
| NGC 720 | -21.2 | C | 0.7 | 2.2 | -2.2 | ... |
| NGC 1052 | -20.4 | ... | irr | 3.0 | -2.26 | ... |
| NGC 1275 | -23.3 | ... | ... | 4.3 | ... | b |
| NGC 1374 | -19.8 | ... | 0.0 | 4.9 | -1.8 | s |
| NGC 1379 | -19.9 | ... | 0.2 | 3.4 | -2.1 | s |
| NGC 1387 | -20.2 | ... | ... | 3.2 | -2.2 | s |
| NGC 1399 | -21.7 | C | 0.1 | 12.4 | -1.6 | b |
| NGC 1404 | -21.0 | ... | 0.5 | 3.5 | -2.0 | s |
| NGC 1427 | -20.0 | ... | 0.7 | 5.1 | -2.0 | s |
| NGC 1549 | -20.8 | C | -0.4 | 0.8 | -1.8 | ... |
| NGC 1553 | -21.0 | ... | ... | 2.3 | -2.3 | ... |
| NGC 3311 | -22.3 | ... | ... | 15.0 | -1.3 | b |
| NGC 3115 | -21.1 | P | ... | 2.3 | -1.84 | ... |
| NGC 3115Dw1 | -17.7 | ... | ... | 4.9 | -1.8 | s |
| NGC 3226 | -19.6 | ... | ... | 7.0 | -2.5 | ... |
| NGC 3377 | -20.1 | P | 1.2 | 2.1 | -1.9 | ... |
| NGC 3379 | -21.0 | C | 0.2 | 1.2 | -1.8 | ... |
| NGC 3384 | -20.3 | P | 0.0 | 1.1 | ... | ... |
| NGC 3557 | -22.6 | ... | 0.0 | 0.4 | ... | ... |
| NGC 3607 | -20.7 | ... | irr | 4.2 | -2.6 | ... |
| NGC 3842 | -23.1 | C | -0.3 | 7.7 | -1.2 | ... |
| NGC 3923 | -22.1 | C | -0.4 | 6.4 | ... | b |
| NGC 4073 | -23.1 | ... | ... | 4.8 | -0.95 | ... |
| NGC 4278 | -19.8 | C | -1.0 | 8.7 | -1.85 | s? |
| NGC 4340 | -20.0 | ... | ... | 8.0 | ... | ... |
| NGC 4374 | -21.7 | C | -0.4 | 6.6 | | s? |
| NGC 4365 | -21.8 | C | -1.1 | 5.0 | -1.15 | b |
| NGC 4406 | -21.8 | C | -0.7 | 6.3 | ... | s? |
| NGC 4472 | -22.6 | C | -0.3 | 5.6 | -1.68 | b |
| NGC 4486 | -22.4 | C | 0.0 | 13.9 | -1.61 | b |
| NGC 4494 | -21.0 | ... | 0.3 | 5.4 | -1.06 | b? |
| NGC 4526 | -21.4 | ... | ... | 7.7 | ... | ... |
| NGC 4552 | -21.2 | C | -2.0 | 8.0 | ... | b? |
| NGC 4564 | -20.1 | P | 2.2 | 10.0 | ... | ... |
| NGC 4621 | -21.2 | P | 1.5 | 6.3 | ... | ... |
| NGC 4636 | -21.7 | C | -0.2 | 7.5 | -1.0 | ... |
| NGC 4649 | -22.2 | C | -0.5 | 6.9 | ... | ... |
| NGC 4697 | -21.6 | P | 1.4 | 2.5 | -1.9 | ... |
| NGC 4874 | -23.0 | C | ... | 14.3 | ... | ... |
| NGC 4881 | -21.6 | ... | ... | 1.0 | 0.0 | ... |

**TABLE 1 (con't.)**

PROPERTIES OF GLOBULAR CLUSTER SYSTEMS HAVING EARLY-TYPE PARENT GALAXIES[a]

| Designation | $M_v$ | Profile type | $100\underline{a}(4)/\underline{a}$ | S | α | Color |
|---|---|---|---|---|---|---|
| NGC 4889 | -23.5 | C | irr | 6.9 | ... | ... |
| NGC 5018 | -22.6 | ... | ... | 1.1 | -1.3 | b |
| NGC 5128 | -22.0 | ... | ... | 2.6 | -1.5 | b |
| NGC 5481 | -20.2 | ... | ... | 2.5 | -1.7 | s? |
| NGC 5629 | -21.7 | ... | ... | 5.0 | ... | ... |
| NGC 5813 | -21.0 | C | irr | 7.2 | -2.18 | s |
| NGC 5846 | -22.1 | ... | 0.0 | 4.5 | ... | ... |
| NGC 6166 | -22.7 | C | ... | 9.0 | -0.95 | s? |
| NGC 7768 | -22.9 | C | ... | 2.8 | -1.3 | ... |
| UGC 9958[c] | -23.4 | ... | ... | 12.0 | -1.2 | ... |
| UGC 9799[d] | -23.4 | C | ... | 21.0 | -1.4 | ... |

[a]     Mostly taken from Kissler-Patig (1997b)
[b]     P = power-law profile, C = profile with central core
[c]     In A 2107
[d]     In A 2052

# TABLE 2

SPECIFIC CLUSTER FREQUENCY AND RADIAL DENSITY GRADIENT OF CLUSTERS

| $\alpha$ | S < 5 | S > 5 |
|---|---|---|
| < -1.75 | 14 | 4 |
| > -1.75 | 8 | 10 |

## TABLE 3

COLOR DISTRIBUTION AND RADIAL DENSITY GRADIENT OF GLOBULAR CLUSTER SYSTEMS

| α | b(road) | s(harp) |
|---|---------|---------|
| < -1.75 | 0 | 7 |
| > -1.75 | 7 | 0 |

# TABLE 4

Chi-squared distributions for 2x2 contingency tables between parameters describing clusters and parent galaxies

|     | C/P | $M_V$ | $\underline{a}(4)$ |
|-----|-----|-------|--------------------|
| S   | 5.2 | 4.1   | 4.7                |
| $\alpha$ | 2.2 | 23 | 3.0            |
| b/s | ... | 16    | 3.9                |

# TABLE 5

RADIAL PROFILE TYPE AND DEVIATIONS FROM ELLIPTICAL ISOPHOTES

| a(4) | P(ower-law) | C(ore) |
|---|---|---|
| + | 4 | 3 |
| − | 0 | 11 |

# TABLE 6

PARENT GALAXY LUMINOSITY AND RADIAL PROFILE TYPE

| $M_V$ | P(ower-law) | C(ore) |
|---|---|---|
| < -21.5 | 1 | 16 |
| > -21.5 | 6 | 6 |

## TABLE 7

### Isophote shape and parent galaxy luminosity

| $M_V$ | $a(4) = +$ | $a(4) = -$ |
|---|---|---|
| < -21.5 | 2 | 8 |
| > -21.5 | 9 | 3 |

## FIGURE CAPTIONS

Fig. 1  Specific globular cluster frequency S as a function of the slope $\alpha$ of their radial density distribution.  The figure shows that high S galaxies tend to have shallow slopes, whereas low S galaxies generally exhibit steep slopes.

Fig. 2  Plot of the isophote shape parameter $\underline{a}(4)/\underline{a}$ versus parent galaxy luminosity $M_V$.  The figure shows that the majority of "disky" early-type galaxies are fainter than $M_V = -21.5$ ($H_o = 75$ km s$^{-1}$ Mpc$^{-1}$ assumed), whereas "boxy" early-type galaxies are mostly brighter than $M_V = -21.5$.

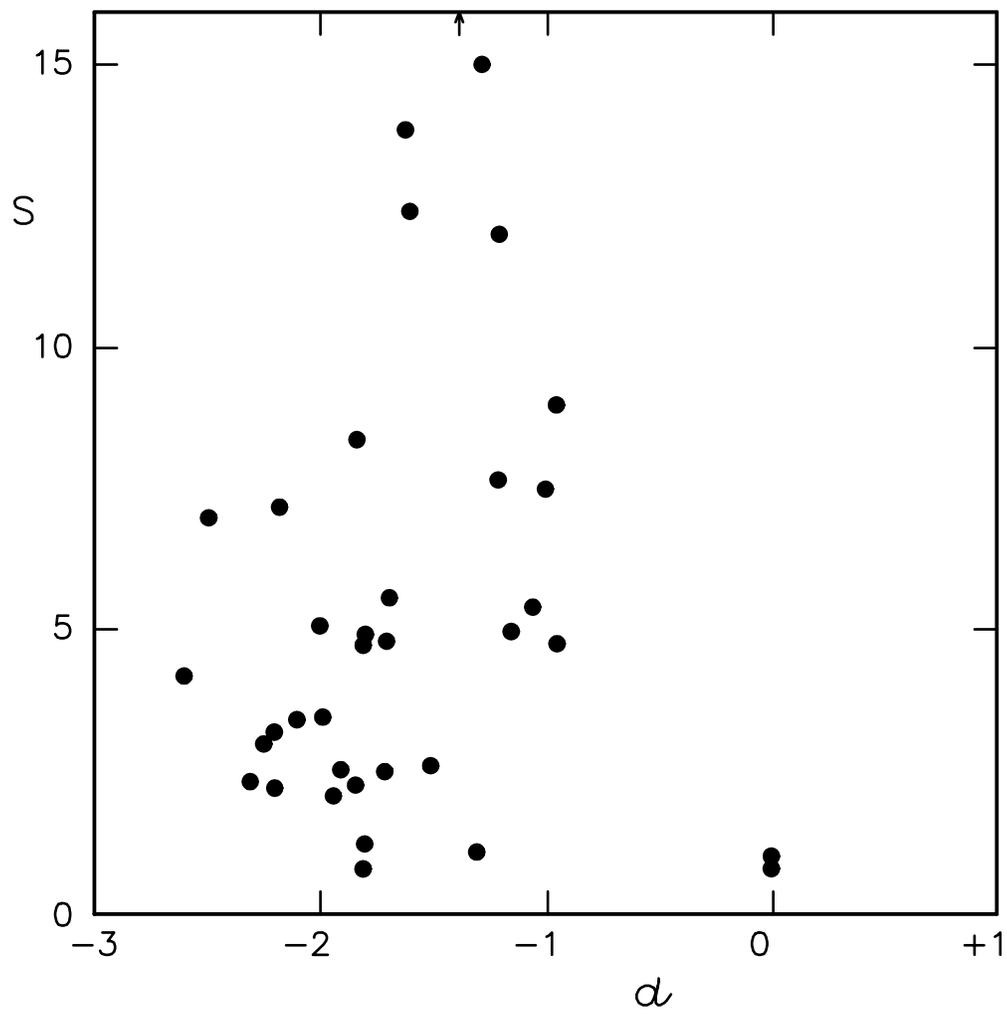